\documentclass[useAMS,usenatbib]{mn2e}
\usepackage{amssymb,amsmath}
\usepackage{graphicx}                                                                                
\title[origination of FeII emission lines from accretion disk]
{SDSS J2125-0813: the evidence for the origination of optical FeII
emission lines from accretion disk near central black hole}
                                                                                
\author[Zhang et al.]
       {Xue-Guang, Zhang$^{1,2}$\thanks{xguang@astroscu.unam.mx},
        Dultzin-Hacyan, D.$^1$,
        Ting-Gui, Wang$^2$ \\
       $^1$Instituto de Astronomia, Universidad Nacional Autonoma de
                 Mexico, Apdo Postal 70-264, Mexico D. F. 04510, Mexico \\
       $^2$Center for Astrophysics, Department of astronomy and Applied
                 Physics, University of Science and Technology of China, \\
                 Hefei, Anhui, P.R.China}

\date{}
\pagerange{\pageref{firstpage}--\pageref{lastpage}} \pubyear{2006}
\def\LaTeX{L\kern-.36em\raise.3ex\hbox{a}\kern-.15em
    T\kern-.1667em\lower.7ex\hbox{E}\kern-.125emX}

\begin{document}
\label{firstpage}

\maketitle

\begin{abstract}
  We report on the radio quiet quasar SDSS J2125-0813  
which obviously emits optical FeII emission lines and double-peaked broad 
Balmer emission lines. Using the accretion disk model for double-peaked broad 
low-ionization emission lines, we reproduce the composite line 
spectra at the optical band between 4100$\AA$ and 5600$\AA$. 
Broad FeII emission lines can be fit simultaneously with
the broad H$\beta$ and MgI lines, such that all broad lines have an elliptical disk profile with the same disk parameters.
This results indicates that the optical FeII emission lines 
originate from the accretion disk near the central black hole which produces 
the double-peaked broad Balmer emission lines. Furthermore, 
we find that the object has dimensionless accretion rate 
$\dot{m}\sim0.4-0.6$ which is much larger than 
accretion rate for ADAF mode, and that the energy budget of the accretion 
disk is enough to power 
the double-peaked broad Balmer emission lines.  
\end{abstract}


\section{Introduction}
    Greenstein \& Schmidt (\citeyear{gs64}) first identified 
FeII emission lines in the spectra of quasars. FeII emission lines have 
been observed and studied in Active Galactic Nuclei (AGN) for more 
than fifty years, and some meaningful results es have 
been found. The ratio FeII4570/H$\beta$ and the equivalent width
W(FeII4570) anti-correlate with the line width of H$\beta$ (Boroson et al.,
\citeyear{bo85}; Zheng \& O'Brien, \citeyear{zo90}; Zheng \& Keel,
\citeyear{zk91}). FeII emission is weak in steep spectrum radio sources 
(Boroson \& Green, \citeyear{bg92}). A large value of the ratio  FeII4570/H$\beta$ corresponds to a high
soft x-ray variability (Moran et al., \citeyear{mo96}). A detailed study of the
line-related correlations for FeII emission lines can be found in 
Sulentic, Marziani \& Dultzin-Hacyan (2000): for example, the strong 
anti-correlation between EW(FeII4570)/EW(H$\beta$) and FWHM($H\beta$). These correlations are discussed in the 
context of the so called  Eigenvector 1  
parameter space where the authors  define a population A 
(FWHM(H$\beta_{broad}$)$<4000km\cdot s^{-1}$), and a population B 
(FWHM(H$\beta_{broad}$)$>4000km\cdot s^{-1}$), which turns out to be a cleaner 
distinction for objects with FeII emission lines than Radio-loud vs. 
Radio quiet objecs.

   It is widely accepted that there are two main kinds of models which 
can reproduce the observed shape and equivalent width of FeII emission lines: 
Photoionization models with microturbulence (Korisra et al., \citeyear{ko97};
Bottorff et al., \citeyear{bo00}; Netzer, \citeyear{ne85}; Baldwin et al.,
\citeyear{bf04}) in the context of LOC model (Baldwin, 1995) and 
collisionally excited models (Grandi, \citeyear{gr81},
\citeyear{gr82}; Kwan et al., \citeyear{kw95}; Dumont et al., \citeyear{du98};
Baldwin et al., \citeyear{bf04}). Due to the large amount of FeII emission 
lines, it is hard to discriminate between photoionization and collisional excitation. 
There are two ways to estimate the intensities of FeII emission lines, 
theoretic line ratios predicted from a model, as is done in  this paper, 
and empirical meassueremnts from FeII templates adapted from I Zwicky 1 
(PG 0050+124) (Philips, 1978; Oke \& Lauer, 1979; Borson \& Green, 1992; 
Laor et al., 1997; Marziani et al., 2003).  

  The origin of FeII emission lines is one of the problems which have not
been solved completely. 
Generally, there are two possible origins: the FeII emission lines comes 
from an illuminated (photoionized) region in the accretion disk near 
central black hole or from shock heated regions in the base of a jet.
Both scenarios depend on the
inclination of the accretion disk which can give an explanation to the 
small number of strong FeII emission line objects, and both scenarios provide the high 
density and high optical depth required for strong FeII emission lines. 
According to the anti-correlation between EW(FeII)/EW(H$\beta$) and 
line width of broad H$\beta$, some authors (Zheng \& Keel, 
\citeyear{zk91}; Joly, \citeyear{jo87}; Wang et al., \citeyear{wz96}; 
Sulentic, Marziani \& Dultzin-Hacyan, 2000) predict that the strength 
of FeII emission lines may be influenced by the inclination 
angle of accretion disk near central BH and by the luminosity-to mass 
ratio of active nucleus. Furthermore, FeII emission lines may be one 
of the important parameters to test different models of AGN (L\'{i}pari et
al., \citeyear{lt93}).

There is a way to inspect whether the FeII emission lines originate from the accretion disk near the central black hole, similarly to the broad lines in a special class of AGN: the double-peaked broad low-ionization emission line 
objects (hereafter, dbp emitters).
The double-peaked broad low-ionization 
emission lines are considered to come from the accretion disk near central 
black hole (Chen \& Halpern, \citeyear{ch89}; Chen et al., 1989; 
Chen et al., \citeyear{ch97}; Eracleous \& Halpern, \citeyear{eh94}; 
Storchi-Bergmann et al., \citeyear{sb03}, Chakrabarti \& Wiita, 1994; 
Hartnoll \& Blackman, 2002; Karas, Martocchia, \& Subr 2001). 
There are some signs that objects with 
very broad Balmer lines actually have weak or absent optical FeII emission 
lines, thus, there are no clues about the properties of optical FeII 
emission lines for dbp emitters, because the line width of braod balmer lines 
is about six times larger than that of normal AGN (Strateva et al., 2003).
Here we report on one quasar, SDSS J212501.21-081328.6 at $z = 0.623$(hereafter, 
SDSS J2125-0813, the redshift is provided by SDSS), selected from SDSS,  
which has both double-peaked broad 
Balmer emission lines, and relatively strong optical and UV FeII emission lines. The 
structure of the paper is as follows: Section 2 gives the analysis of the observed spectrum 
and results. Section 3 gives a discussion and conclusions. The cosmological 
parameters $H_{0}=70{\rm km\cdot s}^{-1}{\rm Mpc}^{-1}$, 
$\Omega_{\Lambda}=0.7$ and $\Omega_{m}=0.3$ have been adopted here.

\section{Observation and Results}
  The apparent optical PSF magnitudes of SDSS J2125-0813 are 17.78, 17.13, 
17.07, 16.96, 17.01 at u, g, r, i and z band respectively. The radio flux at
20cm of SDSS J2125-0813 is 1.59mJy provided by FIRST. The radio-loudness 
($R=L_{20cm}/L_{4400\AA}$),  
2.95, indicates SDSS J2125-0813 is a radio-quiet quasar. Using the 
SDSS 2.5m telescope sited at Apache Point Observatory, the
reddening corrected (with E(B-V) = 0.068) spectrum of SDSS J2125-0813 with
total exposure time 8406 seconds is shown in Figure 1.

  Before proceeding further, a careful analysis of the continuum is 
carried out. From the spectrum, there are obviously double-peaked H$\beta$, 
H$\gamma$ and an 3100$\AA$ bump which is made up of UV FeII emission lines 
and Balmer
continuum emission (Neugebauer et al., \citeyear{ng87}). 
Thus when we fit the continuum, we avoid these features. 
We select three 
 wavelength regions, from 2900$\AA$ to 3050$\AA$, from 3950$\AA$ to 
4150$\AA$ and from 5400$\AA$ to 5600$\AA$ to determine the continuum. 
The fit results with the format $f_{\lambda}\propto\lambda^{-1.99}$ 
are shown in Figure 1. SDSS 
J2125-0813 is a flat-spectrum radio-quiet quasar. We notice that 
there is a much weaker double-peaked H$\delta$ around 4100$\AA$. So the 
level of the continuum might be a little overestimated. The 
spectrum at optical band after the subtraction of continuum is also shown in 
Figure 1. Furthermore, we select the wavelength range from $\sim$4950$\AA$ 
to $\sim$5050$\AA$ to determine the line parameters of the [OIII] doublet by  
two gaussian functions adding a backgroud. According
to the fit results, the line width of the [OIII] doublet is the same
within the error, and the line flux ratio is near to 0.33 within the error. The
best fit results for the [OIII] doublet are shown in Figure 1. 
We notice that the center wavelength of [OIII]$\lambda5007\AA$, $5014\AA$,
deviates from the vacuum wavelength (5008.24$\AA$), the same deviation of the center 
wavelength is found for [OII]$\lambda3727\AA$. Thus, the redshift of 
SDSS J2125-0813 should be corrected to 0.621.

  After the subtraction of continuum, we first try to fit the 
double-peaked H$\beta$ using the elliptical accretion disk model 
(Eracleous et al., \citeyear{er95}). But, we notice the presence of obvious
FeII and/or MgI and/or FeIII emission lines sitting on the wings of H$\beta$ 
(The wings of the double-peaked broad emission lines depend sensitively on 
inner radius of accretion disk model).
We do not get the complete line profile of double-peaked H$\beta$, so we fit 
all the line spectra between 4100$\AA$ and 5600$\AA$. In this range of  
wavelength, there are H$\beta$, H$\gamma$, [OIII]$\lambda4959,5007\AA$, 
FeII emission lines, MgI and/or FeIII emission lines. 
The [OIII] doublet can be subtracted using the results above. 
The double-peaked line profile of H$\beta$ in velocity space, 
$f_{H\beta}$, can be reproduced 
by the elliptical disk model with 8 free parameters. The line profile of the 
double-peaked H$\gamma$ in velocity space is estimated by 
$k_{H\gamma/H\beta} * f_{H\beta}$, 
i.e., the line profile of H$\gamma$ is the same as that of H$\beta$ in 
velocity space, 
except for the flux density, where $k_{H\gamma/H\beta}$ is the flux ratio 
of H$\gamma$ to H$\beta$. Furthermore, we assume that FeII 
emission lines in SDSS J2125-0813 are also broadened by the orbiting 
elliptical accretion disk, and have the same double peaked line profile as that of 
H$\beta$ in velocity space. The same assumptions are valid for MgI and FeIII 
emission lines. The line profiles in velocity space of 
MgI and optical FeII, FeIII 
emission lines are estimated using: $k_{MgI_{tot}/H\beta} * f_{H\beta}$,
$k_{FeII_{tot}/H\beta} * f_{H\beta}$ and  $k_{FeIII_{tot}/H\beta} * f_{H\beta}$
Thus there are 12 free parameters in the fitting procedure.

  Theoretical FeII emission line strengths for
physical conditions typical of AGN with BLR have been presented by Sigut \&
Pradhan (\citeyear{sp03}) who considered the following excitation mechanisms: 
continuum fluorescence (Phillips, \citeyear{ph78}, \citeyear{ph79}), 
collisional excitation (Joly, \citeyear{jo91}), self-fluorescence amoung 
the FeII transitions and fluorescent excitation by Ly$\alpha$ and Ly$\beta$ 
(Penston, \citeyear{pe88}; Sigut \& Pradhan, \citeyear{sp98}; Verner et al.,
\citeyear{ver99}). We collect all 36 FeII emission lines in the 
wavelength range 
from 4100$\AA$ to 5600$\AA$ from  Sigut \& Pradhan 
(\citeyear{sp03}). In our fitting procedure, the flux ratio of different FeII
emission lines is the same as the theoretical value. For MgI emission lines, we
select the three strongest lines between 5100$\AA$ and 5200$\AA$, MgI5185,
5174, 5168$\AA$, the theoretical line strength ratios are 1:1.66:4.98. For 
optical FeIII emission lines, we collect the 26 strongest lines between 
5000$\AA$ and 5600$\AA$ (the optical FeIII emission lines between 4100$\AA$ 
and 5000$\AA$ are much weaker). The theoretical line strength ratios are 
selected from the website of Atomic Line List 
(http://www.pa.uky.edu/~peter/atomic/).

  Using the Levenberg-Marquardt technique and ajusting the 12 free parameters 
(MPFIT package in IDL),  
we can get the best results of the 12 parameters
for the line spectra after the subtraction of [OIII] doublet and 
continuum. Last, we accept the flux ratio of H$\beta$ to H$\gamma$ as 3.7,
which indicates some intrinsic absorption. The flux ratio of H$\beta$ to 
total optical FeII emission lines is 1.85, and the ratio of H$\beta$ to 
total MgI emission lines is 87.51. There is almost no optical FeIII emission 
lines. The best fit results are not good near 4600$\AA$, we
think this is due to CVI$\lambda4607\AA$, NIV$\lambda4608\AA$ and
ArII$\lambda4610\AA$ emission lines.
The last best fit results are shown in Figure 2. The value of the 
added squared residuals is 1996 with degrees of freedom 1351, 
$\chi^{2}=1996/1351=1.47$.
The inner radius of the elliptical disk is about 31$\pm6R_G$, the outer 
radius is about 339$\pm40R_G$. The eccentricity of the disk is nearly zero. 
The slope of emissivity is about $\epsilon\propto r^{-1.58\pm0.19}$. The 
inclination angle is about $15.2\pm1.6$ degrees. The local broadening 
velocity $\sigma$ is about 2617$\pm277km\cdot s^{-1}$. The excellent 
fit to the FeII emission lines supports the conclusion that the optical FeII 
emission lines are coming from the same place in the disk as the 
double-peaked broad H$\beta$.

   We examine the possibility that the FeII emission lines 
originate from "normal" Broad Line clouds.  We try to fit 
the pure FeII spectrum after subtraction of the Balmer emission, the [OIII] 
doublet, etc. using Gaussian profiles. The result yields a FWHM for  
each FeII emission line of about 14998$\pm605km\cdot s^{-1}$ with 
$\chi^{2}\sim1.68$. This is an unreasonable value for normal BLR clouds. 
Furthermore, if we notice that the FWHM of H$\beta$ is also
about 14800$km\cdot s^{-1}$, the coincidence in velocities indicates that FeII 
emission lines originate from the same place as the broad H$\beta$. According 
to the value of $chi^2$ and the same line width, we think the accretion 
disk model gives a better fit.

\section{Discussion and Conclusions}

  SDSS J2125-0813 has been mentioned as one of dbp emitters in 
other papers (Wu \& Liu, 2004; 
Strateva, et al., 2003), however, there is no discussion about 
the mass of the BH. In this case the continuum luminosity and line width 
of braod H$\beta$ can not be used to estimate the mass of central black hole under the 
assumption of virialization. The size of broad emission line 
regions (BLRs) estimated 
by the empirical relation  $R_{BLR} - L_{5100\AA}$ (Wandel et al., 1999; 
Kaspi et al., 2000; 2005; Peterson et al., 2004), 
$R_{BLR}\sim234.5 light-days$, is much larger than the size of the 
accretion disk (for $M_{BH}\sim10^{8}M_{\odot}$, $1000R_{G} \sim 5.6$ 
light-days). We will disscuss the difference 
between the true size of BLRs of dbp emitters and the size of BLRs 
from the empirical relation in a forthcoming paper.
In this case, we estimate the mass of the BH with another method through 
the line width of the [OIII] emission lines as tracers of the stellar 
velocity dispersion $\sigma$ (Tremaine, et al., 2002; Gebhardt et al., 
2000; Nelson \& Whittle, 1996; Boroson, 2003), which yields a value of
$M_{BH}\sim 10^{8.13}(\sigma/200)^{4.02}\sim3.1_{-0.5}^{+0.7}\times10^{8}M_{\odot}$. If the mass of the BH is estimated with  from the continuum luminosity (Peterson et al., 2004),
$M_{BH}\propto \lambda L_{5100\AA}^{0.79} \sim 8.4_{-1.8}^{2.8}10^7M_{\odot}$,
the value is not very different from that estimated by the line width of
[OIII]$\lambda5007\AA$.
The mass of the BH confirms that the size of the accretion disk is much 
smaller than the radius estimated from the continuum luminosity. 

  We can also estimate the dimensionless accretion rate, 
$\dot{m}= L_{bol}/L_{Edd}$, where $L_{Edd}$ is the Eddington luminosity given 
by $L_{Edd}=1.26\times10^{46}erg\cdot s^{-1}M_{BH}/10^8M_{\odot}$ and 
$L_{bol}$ is the bolometric luminosity given by 
$L_{bol}\sim9 - 13\times L_{5100\AA}$ (Wandel et al., 1999;  Kaspi et al., 
2000). The accretion rate is $\dot{m}\sim0.4 - 0.6$, much larger than the mean 
value of accretion rate 0.01 for the sample of double-peaked emmiters 
in the paper of Wu \& Liu (2004). In their paper, 
the largest value of accretion rate is about 0.12 for SDSS J1710+6521.  This is in agreement with the tendency for strong FeII
emmiters to be found in large accretion rate (population A according to Sulentic, Marziani \& Dultzin-Hacyan, 2000) objects.
We also find that the optical-X-ray spectral index is $\alpha_{OX}=1.2$ for 
SDSS J2125-0813 with higher luminosity 
$L_{5100\AA}\sim10^{45.36}erg\cdot s^{-1}$,  which 
is not larger than the mean value for low luminosity double-peaked broad line 
objects found in nearby galaxies, $\alpha_{OX}\sim0.9 - 1.1$.  

   We can also estimate the energy budget of the disk by integrating the 
formula for standard thin accretion disk mode, 
$W_{disk}(r_{in},r_{out})\sim2.58\times10^{43}\times\eta^{-1}erg\cdot s^{-1}$, 
where $\eta$ is efficiency of conversion of energy. The luminosity of 
H$\beta$ is about $L_{H\beta}\sim3.14\times10^{43}erg\cdot s^{-1}$.  
If we accept $\eta\sim0.01-0.06$, then the energy budget of the accretion disk 
could be enough to power the broad double-peaked Balmer line. 
When the accretion disk model was first proposed for double-peaked broad 
low-ionization emission line object Arp 102B (Chen et al., 1989), the 
luminosity of H$\alpha$ exceeded the power output of the accretion disk. 
Thus, an ion torus or a hot corona around the inner accretion disk was 
definitely needed for illumination on the regions which produce double-peaked 
emission lines from the inner part of the accretion disk. This problem 
of energy budget is confirmed in other double-peaked  
objects which are clssified as LINERs in the paper of Eracleous \& Halpern 
(1994). If there is an ion torus in the inner part of the accretion 
disk, the blue bump should be much weaker. However, SDSS J2125-0813 is a 
normal quasar with an obvious UV bump, thus the ion torus in the inner part of 
the accretion disk is not necessary. The normal quasar spectra 
also indicates that the accretion mode is not ADAF which is consistent with 
the value of dimensionless accretion rate.

  The object studied in this paper obeys the strong correlation between 
EW(FeII4570)/EW(H$\beta$) and FWHM(H$\beta$) (Sulentic, Marziani \& 
Dultzin-Hacyan, 2000). The result 
confirms that Population A/B in PCA Eigenvector 1 space is a cleaner
distinction for strong FeII emission lines. Furthermore, for radio loud 
object, the continuum should consist of two components, one from nucleus, 
another from jet, which leads to a smaller value of EW(FeII4570). For 
radio loud dbp emitters, the contribution of jet is much larger 
because of inclination angle. Whether the FeII properties of other dbp emitters 
are the same as that of SDSS J2125-0813 will be studied in a following paper.
  
  The seperation of the blue peak ($\sim4840\AA$) from the 
center wavelength of H$\beta$, $\sim20\AA$, is much smaller than 
that of the red peak ($\sim4960\AA$), $\sim100\AA$. This does not obey the 
correlation between peak sepeartion and dimensionless accretion rate found in
Wu \& Liu (2004). This 
is perhaps due to the different way to estimate BH mass which leads to 
a higher dimensionless accretion rate. The line profile best fit 
by an elliptical disk model with eccentricity $e\sim 0$ could reflect the 
true physical situation to some extend. The line profile of FeII emission 
 fit by the same  profile as H$\beta$ indicates that the 
FeII emission line regions are the same as those of H$\beta$  
in SDSS J2125-0813. There is another model for 
double-peaked broad emission lines, 
a double-stream along radio jet model (Zheng, Binetie \& Sulentic, 1990). 
Their results imply that the blue peak should always 
be higher than the red peak, the blue peak seperation should always 
be larger than the red peak seperation, if the physical environment is 
the same for both sides of accretion disk. Furthermore, variability 
studies of double-peaked emission lines favor the accretion disk 
model (Newman et al., 1997; Gilbert et al., 1999; 
Storchi-Bergmann et al., 2003), 
and reverberation mapping experiments have ruled out a bipolar outflow 
in the canonical dbp emitter 3C 390.3 (Dietrich et al., 1998).

\section*{Acknowledgments}
We gratefully thank the referee's comments that allowed us to improve
and clarify our manuscript.
ZXG gratefully acknowleges the postdoctoral scholarships offered by la 
Universidad Nacional Autonoma de Mexico (UNAM). D. D-H acknowledges 
support from grant IN100703 from DGAPA, UNAM.
This paper has made use of the data from the SDSS projects. 
Funding for the creation and the distribution of the SDSS Archive
has been provided by the Alfred P. Sloan Foundation, the
Participating Institutions, the National Aeronautics and
Space Administration, the National Science Foundation,
the U.S. Department of Energy, the Japanese Monbukagakusho, and the 
Max Planck Society. The SDSS is managed by the Astrophysical Research 
Consortium (ARC) for the Participating Institutions. The Participating 
Institutions are The University of Chicago, Fermilab, the 
Institute for Advanced Study, the Japan Participation Group,
The Johns Hopkins University, Los Alamos National Laboratory, 
the Max-Planck-Institute for Astronomy (MPIA),
the Max-Planck-Institute for Astrophysics (MPA), New
Mexico State University, Princeton University, the United
States Naval Observatory, and the University of Washington.

\begin{figure}
\centering\includegraphics[width = 84mm]{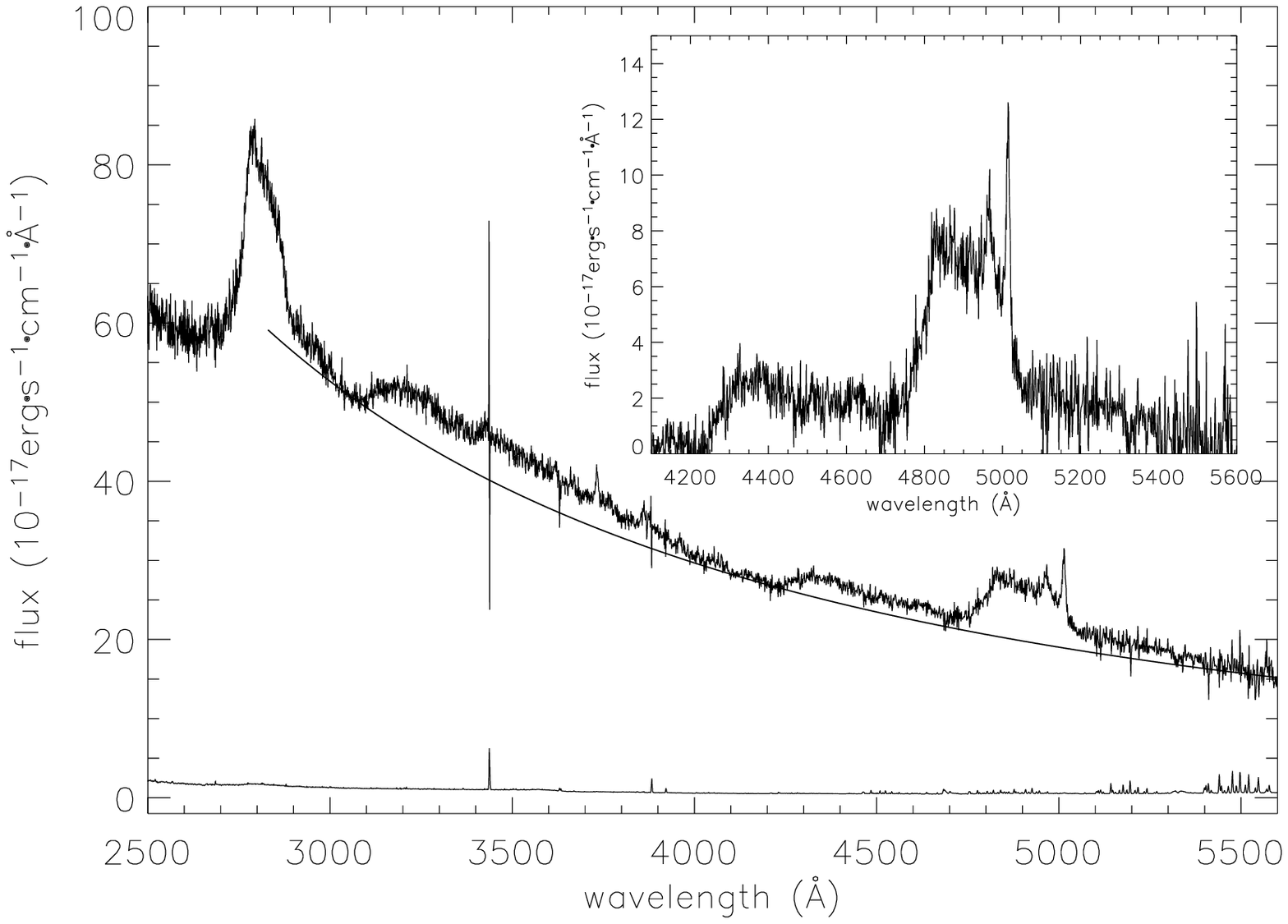}
\centering\includegraphics[width = 84mm]{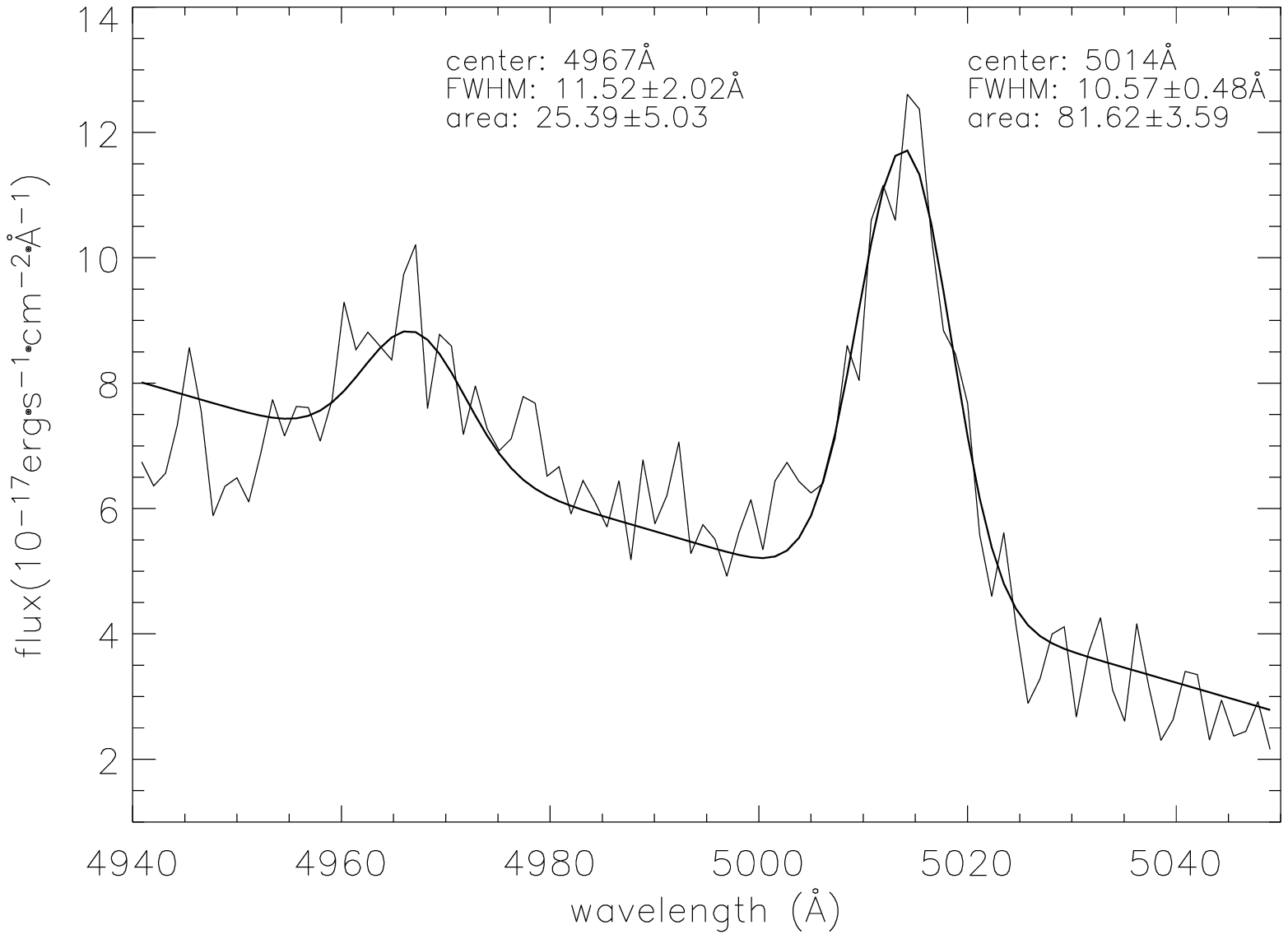}
\caption{The spectrum of SDSS J2125-0813 in rest wavelength is shown in upper
pannel. The continuum is shown in the figure as solid line. 
The spectrum between 4100$\AA$ and 5600$\AA$ is shown in the 
upper-right pannel. The bottom pannel shows the fit results of [OIII]
doublet. The fit parameters are listed in the figure.}
\end{figure}

\begin{figure}
\centering\includegraphics[width = 84mm]{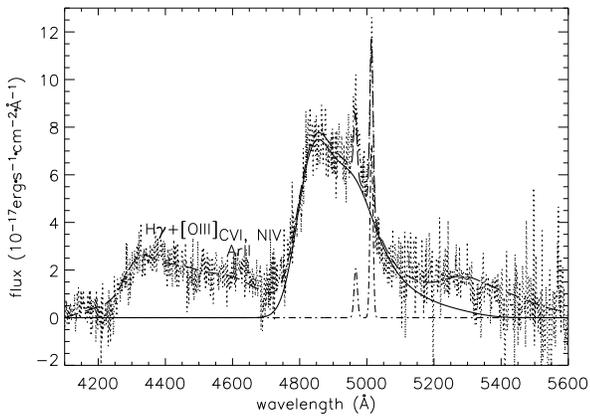}
\caption{The best fit results for spectra smoothed by 6 points 
between 4100$\AA$ and 5600$\AA$ are shown in the figure. 
The dotted line represents the observed spectra. The dashed 
line represents the best fit results. The solid line
represents the profile of H$\beta$. The dash dot line represents the 
[OIII] doublet.
}
\end{figure}                                                                    
\label{lastpage}

\begin{thebibliography}{}
\bibitem[Baldwin et al., 1995]{bf95}
Baldwin, J. A., Ferland, G. F., Korista \& Verner, D., 1995, ApJ, 455, L119
\bibitem[Baldwin et al., 2004]{bf04}
Baldwin, J. A., Ferland, G. F., Korista, K. T., et al., astro-ph/0407404
\bibitem[Boroson et al., 1985]{bo85}
Boroson, T., A., Persson, S. E. \& Oke, J. B., 1985, ApJ, 293, 120
\bibitem[Boroson \& Green, 1992]{bg92}
Boroson, T., A. \& Green, R. F., 1992, ApJS, 80, 109
\bibitem[Boroson, 2003]{bo03}
Boroson, T. A., 2003, ApJ, 585, 647
\bibitem[Bottorff et al., 2000]{bo00}
Bottorff, M., Ferland, G., Baldwin, J. \& Korista, K., 2000, ApJ, 542, 644
\bibitem[Chakrabarti \& Wiita, 1994]{cw94}
Chakrabarti, S. K. \& Wiita, P. J. 1994, ApJ, 434, 518
\bibitem[Chen \& Halpern, 1989]{ch89}
Chen, Kaiyou \& Halpern, J. P., 1989, APJ, 344, 115
\bibitem[Chen et al., 1989]{c89}
Chen, Kaiyou, Halpern, J. P. \& Filippenko, A .V., 1989, ApJ, 339, 742
\bibitem[Chen et al., 1997]{ch97}
Chen, Kaiyou, Halpern, J. P. \& Titarchuk, L. G., 1997, ApJ, 483, 194
\bibitem[Dietrich et al., 1998]{di98}
Dietrich, M., et al., 1998, ApJS, 115, 815
\bibitem[Dumont et al., 1998]{du98}
Dumant, A. M., Collin-Souffrin, S. \& Nazaiova, L., 1998, A\&A, 331, 11
\bibitem[Eracleous \& Halpern, 1994]{eh94}
Eracleous, M. \& Halpern, J. P., 1994, ApJS, 90, 1
\bibitem[Eracleous et al., 1995]{er95}
Eracleous, M., Livio, M., Halpern, J. P. \& Storchi-Bergmann, T., 1995, ApJ,
438, 610
\bibitem[Gebhardt et al., 2000]{gbb00}
Gebhardt, K., Bender, R., Bower, G., et al., 2000, ApJ, 539, L13
\bibitem[]{}
Gilbert, A. M., Eracleous, M., Filippenko, A. V. \& Halpern, J. P., 1999, 
ASPC, 175, 189
\bibitem[Grandi, 1981]{gr81}
Grandi, S. A., 1981, ApJ, 251, 451
\bibitem[Grandi, 1982]{gr82}
Grandi, S. A., 1982, ApJ, 255, 25
\bibitem[Greenstein \& Schmidt, 1964]{gs64}
Greenstein, J. L. \& Schmidt, M., 1964, ApJ, 140, 1
\bibitem[Hartnoll \& Blackman, 2002]{hb02}
Hartnoll, S. A. \& Blackman, E. G. 2002, MNRAS, 332, L1
\bibitem[Karas, et al., 2001]{kms01}
Karas, V., Martocchia, A. \& Subr, L. 2001, PASJ, 53, 189
\bibitem[Joly, 1987]{jo87}
Joly, M., 1987, A\&A, 184, 33
\bibitem[Joly, 1991]{jo91}
Joly, M., 1991, A\&A, 242, 49
\bibitem[Kaspi et al., 2000]{ka00}
Kaspi, S., Smith, P. S., Netzer, H., et al., 2000, ApJ, 533, 631
\bibitem[Kaspi et al., 2005]{ka05}
Kaspi, S.,Maoz, D., Netzer, H. \& Peterson, B. M., 2005, ApJ, 629, 1
\bibitem[Korista et al., 1997]{ko97}
Korista, K., Baldwin, J., Ferland, G. \& Verner, D., 1997, ApJS, 108, 401
\bibitem[Kwan et al., 1995]{kw95}
Kwan, J., Cheng, F-Z., Fang, L-Z., Zheng, Wei \& Ge, J., 1995, ApJ, 440, 628
\bibitem[Laor et al., 1997]{la97}
Laor, A., Jannuzi, B. T., Green, R. F. \& Boroson, T. A., 1997, ApJ, 489, 656
\bibitem[L\,{i}pari et al., 1993]{lt93}
L\'{i}pari, S., Terlevich, R. \& Macchetto, F., 1993, ApJ, 406, 451
\bibitem[Marziani et al., 2003]{ma03}
Marziani, P., Sulentic, J. W., Zamanov, R., Calvani, M., Dultzin-Hacyan, D., 
Bachev, R., Zwitter, T., 2003, ApJS, 145, 199
\bibitem[Moran et al., 1996]{mo96}
Moran, E. C., Halpern, J. P. \& Helfand, D. J., 1996, ApJS, 106, 341
\bibitem[Nelson \& Whittle, 1996]{nw96}
Nelson, C. H. \& Whittle, M., 1996, ApJ, 465, 96
\bibitem[Netzer, 1985]{ne85}
Netzer, H., 1985, ApJ, 289, 451
\bibitem[Netzer \& Peterson, 1997]{np97}
Netzer, H. \& Peterson, B. M., in Astronomical Time Series, ed. Maoz, D., 
Sternberg, A. \& Leibowitz, E., 85
\bibitem[Neugebauer et al., 1987]{ng87}
Neugebauer, G., Green, R. F., Matthews, K., et al., 1987, ApJS, 63, 615
\bibitem[]{}
Newman, J. A., Eracleous, M., Filippenko, A. V. \& Halpern, J. P., 1997, ApJ, 485, 570
\bibitem[Oke \& Lauer, 1979]{ol79}
Oke, J. B. \& Lauer, T.R., 1979, ApJ, 230, 360
\bibitem[Penston, 1988]{pe88}
Penston, M. V., 1988, MNRAS, 233, 601
\bibitem[Peterson, 1993]{pe93}
Peterson, B. M., 1993, PASP, 105, 247
\bibitem[Peterson et al., 2004]{pe04}
Peterson, B. M., Ferrarese, L., Gilbert, K. M., et al., 2004, ApJ, 613, 682
\bibitem[Phillips, 1978]{ph78}
Phillips, M. M., 1978, ApJ, 226, 736
\bibitem[Phillips, 1979]{ph79}
Phillips, M. M., 1979, ApJS, 39, 377
\bibitem[Sigut \& Pradhan, 1998]{sp98}
Sigut, T. A. A. \& Pradhan, A. K., 1998, ApJL, 499, 139
\bibitem[Sigut \& Pradhan, 2003]{sp03}
Sigut, T. A. A. \& Pradhan, A. K., 2003, ApJS, 145, 15
\bibitem[Storchi-Bergmann et al., 2003]{sb03}
Storchi-Bergmann, T., Nemmen da Silva, R., Eracleous, M., et al., 2003, ApJ,
598, 956
\bibitem[Strateva et al., 2003]{ssh03}
Strateva, Iskra V., Strauss, M. A., Hao L, et al., 2003, AJ, 126, 1720
\bibitem[Sulentic et al., 2000]{smd00}
Sulentic, J. W., Marziani, P. \& Dultzin-Hacyan, D., 2000, ARA\&A, 38, 521
\bibitem[Tremaine et al., 2002]{tgb02}
Tremaine, S., Gebhardt, K., Bender, R., ET AL., 2002, ApJ, 574, 740
\bibitem[Verner et al., 1999]{ver99}
Verner, E. M., Verner, D. A., Korista, K. T., et al., 1999, ApJS, 120, 101
\bibitem[Wandel et al., 1999]{wan99}
Wandel, A., Peterson, B. M. \& Malkan, M. A., 1999, ApJ, 526, 579
\bibitem[Wang et al., 1996]{wz96}
Wang, T-G., Zhou, Y-Y. \& Gao, A-S., 1996, ApJ, 457, 111
\bibitem[Wu \& Liu, 2004]{wl04}
Wu, X-B \& Liu, F-K, 2004, ApJ, 614, 91
\bibitem[]{zb90}
Zheng, Wei, Binetie, L. \& Sulentic, J. W., 1990, ApJ, 365, 115
\bibitem[Zheng \& O'Brien, 1990]{zo90}
Zheng Wei \& O'Brien, P. T., 1990, ApJ, 353, 433 
\bibitem[Zheng \& Keel, 1991]{zk91}
Zheng Wei \& Keel, W. C., 1991, ApJ, 382, 121
\end{thebibliography}
\end{document}